\documentclass[twoside,fleqn]{article}
\usepackage{espcrc2}
\usepackage{psfig}
\newcommand{\beq}{\begin{equation}}
\newcommand{\eeq}{\end{equation}}
\newcommand{\beqn}{\begin{eqnarray}}
\newcommand{\eeqn}{\end{eqnarray}}
\newcommand{\bea}[1]{\beq\begin{array}{#1}}
\newcommand{\eea}{\end{array}\eeq}

\newcommand{\ket}[1]{|\,#1\,\rangle}
\newcommand{\bra}[1]{\langle\,#1\,|}

\newcommand{\diff}{\partial}

\newcommand{\NP}[3]{{\it Nucl. Phys. }{\bf #1} (#2) #3}

\newcommand{\PL}[3]{{\it Phys. Lett. }{\bf #1} (#2) #3}

\newcommand{\CMP}[3]{{\it Comm. Math. Phys. }{\bf #1} (#2) #3}

\newcommand{\PTPS}[3]{{\it Prog. Theor. Phys. Suppl. }{\bf #1} (#2) #3}

\title{The Berry Phase and Monopoles in Gluodynamics\thanks{Presented by F. V. G. at 
Lattice'2001, Berlin.}}

\author{ F.V.~Gubarev~$^{a,b}$, V.I.~Zakharov~$^b$\\[2mm]
		$^{\rm a}$ {\small\it Institute of Theoretical and  Experimental Physics, B.Cheremushkinskaya 25, Moscow, 117259, Russia}\\
		$^{\rm b}$ {\small\it Max-Planck Institut f\"ur Physik, F\"ohringer Ring 6, 80805 M\"unchen, Germany}}
\begin{document}

\begin{abstract}
We introduce a gauge invariant definition of a monopole on the lattice. The construction is based
on the observation that for each Wilson loop there exists an extra $U(1)$ group which leaves the
loop invariant. Since the lattice formulation utilizes the language of Wilson loops, the definition
of the monopole charge in terms of this plaquette dependent $U(1)$ is gauge invariant.
The explicit construction of gauge invariant monopoles is presented both in continuum and on the lattice.
\end{abstract}

\maketitle

Introduction of magnetic monopoles in non-Abelian pure gauge models has a long history
(see \cite{reviews} and references therein). The basic problem is that the monopoles
are intrinsically $U(1)$ objects and, in absence of the physical Higgs field, the choice
of the $U(1)$ turns out to be a matter of gauge fixing.
The standard procedure\footnote{
Throughout this paper we consider the case of $SU(2)$ gauge group only.
} to define monopole charge is to partially (up to a remaining $U(1)$) fix the gauge and then
construct  Abelian monopoles in this particular gauge. Due to the ambiguity of gauge fixing
prescription the construction is not unique and it is a separate question which monopole definition
is physically relevant.

{\bf 1.} The crucial observation \cite{main} which is used in considerations below
is that in lattice formulation there is in fact a natural $U(1)$ subgroup
locally embedded into $SU(2)$. Indeed, the action of $SU(2)$ lattice gluodynamics
is constructed in terms of elementary Wilson loops
$U_p = e^{i \vec{F}_p \vec{\sigma}/2} = e^{i |F_p| \vec{n}_p \vec{\sigma}/2}$,
$\vec{n}^2_p = 1$ and therefore possesses an additional symmetry
$\vec{F}_p \to \vec{F}_p + 4\pi \vec{n}_p$ which is evidently lost in the naive continuum limit.
In other words Wilson loop itself defines a natural $U(1)$ associated with
it as the group of rotations around $\vec{n}_p$. Since these Abelian rotations are defined in terms
of a Wilson loop, the definition of the $U(1)$ is gauge invariant by construction. The application
of this procedure to each plaquette gives, basically, a gauge invariant fixation of $U(1)$ which is
plaquette dependent.

To make the next step and define monopoles in terms of this local, gauge invariant $U(1)$
we utilize the interpretation  of Wilson loop
as an evolution operator of quantum mechanical system \cite{BerryPhase}. As follows from the standard representation
$W(C)=W(T)={\mathrm P} \exp\{i \oint_{\,0}^T A dt\}$ the Hamiltonian governing evolution is $H=-A$
and quantum mechanical state space coincides with irreducible representation space of $SU(2)$.
Here $C$ is an arbitrary closed contour parameterized by $t\in [0;T]$.
Furthermore, we introduce the conventional \cite{perelomov}
coherent state basis $\{\ket{\vec{n}}\}$ parameterized by a set of unit three-dimensional vectors $\vec{n}$,
$\vec{n}^2 = 1$. The action of an arbitrary element $g \in SU(2)$ in this basis is\footnote{
For simplicity we consider fundamental representation only, generalization
to higher representations is straightforward.
} $g \, \ket{\vec{n}} = e^{i\varphi} \, \ket{\vec{n}'}$. A special case of cyclic evolution
\beq
\label{some}
W(T) \, \ket{\vec{n}(0)} = e^{i\varphi(T)} \, \ket{\vec{n}(0)}
\eeq
is of particular importance. In Eq.~(\ref{some}) the evolving state $\ket{\vec{n}(t)}$ is determined by
\beq\label{evolving}
\mathrm{P}\exp\{i\int_0^t A d\tau\} \, \ket{\vec{n}(0)} = e^{i\varphi(t)} \,\ket{\vec{n}(t)}\,.
\eeq
Clearly the cyclic state $\ket{\vec{n}}$ always exists and moreover $\frac{1}{2}\mathrm{Tr}W(T) = \cos\varphi(T)$.
From (\ref{evolving}) it directly follows that 
\beq
\label{phase}
\varphi(T) = \int_0^T\left(\bra{\vec{n}} A \ket{\vec{n}} + i \bra{\vec{n}}\frac{d}{dt}\ket{\vec{n}}\right)dt=
\eeq
$$
\qquad
= \int_C \vec{n}\vec{A} + \frac{1}{2} \int_{S_C} \vec{n} \cdot [ \partial \vec{n} \times \partial\vec{n}]\,,
$$
where the state $\ket{\vec{n}}$ has been smoothly extended from the contour $C$ into an arbitrary surface $S_C$
bounded by $C$.  Note that  Eq.~(\ref{phase}) cannot be used in the Wilson loop calculation
since construction of the evolving state $\ket{\vec{n}}$ requires knowledge of the Wilson loop itself.
Nevertheless, Eq.~(\ref{phase}) is useful theoretically
since it represents the phase angle of Wilson loop as an integral of abelian 't~Hooft tensor \cite{hooft}.
Note also that the angle $\varphi(T)$ is well defined only modulo $2\pi$.
Indeed, one can verify explicitly that $\varphi$ is changing $\varphi \to \varphi + 2\pi k$, $k \in Z$
under gauge transformations. From now on we always take $\varphi(T)$ in ``fundamental domain''
$-\pi < \varphi \le \pi$.

Consider now the infinitesimal version of Eq.~(\ref{phase}) when contour $C$ is the boundary of
elementary surface element $\delta\sigma$
\beq
\label{small}
\varphi = \left( \diff\wedge (\vec{n}\vec{A}) + 
\vec{n} \wedge \diff\vec{n} \wedge \diff\vec{n} \right)\delta\sigma\,.
\eeq
It is straightforward now to integrate (\ref{small}) over arbitrary closed two-dimensional surface
$S^2_{phys}$ in physical space. Usual assumption of fields 
continuity guarantees that $\vec{n}$ is a smooth field on $S^2_{phys}$. Therefore integral
is non-zero in general
\beq
\label{charge}
\int_{S^2_{phys}} \varphi ~=~ 2\pi Q\,, \qquad Q\in Z
\eeq
due to the second term in Eq.~(\ref{small}) which is widely known as Berry phase \cite{Berry}.
In fact Eq.~(\ref{charge}) calculates the gauge invariant monopole charge $Q$ contained inside $S^2_{phys}$.
Gauge invariance is evident since integral is
constructed in terms of infinitesimal Wilson loops only. The term ``monopole charge'' is also
quite natural because Eq.~(\ref{charge}) coincides with well known Abelian monopole construction
when gauge fields have the same color structure.
Moreover, Eq.~(\ref{charge}) being considered in four-dimensional space-time
defines a closed world-lines of topological defects which have a natural interpretation of monopole
trajectories.

To conclude this section we would like to emphasize that gauge invariant monopole charge definition is
only possible when gluodynamics is considered as limiting case of lattice gauge models because
of periodicity of gauge fields action on the lattice (or more details see \cite{main,appear}).
One can verify that  Eq.~(\ref{charge}) considered in the context of conventional continuum
gluodynamics is trivial and produces  identically zero magnetic charge.

{\bf 2.} Being extremely simple in continuum limit, Eq.~(\ref{charge}) is quite non-trivial
to implement on the coarse lattice. The problem is that lattice discretization is not suitable to
calculate most of topological invariants (e.g. instanton number). In our case the problem is even more
severe since effective Higgs field in Eq.~(\ref{small},\ref{charge}) is defined not in lattice sites as usual,
but on elementary 2-cells (plaquettes). Below we illustrate the monopole charge calculation in case
of single three-dimensional cube. More realistic calculations will be presented elsewhere~\cite{appear}.

To begin with we consider a single plaquette situated at lattice site $x$ and directed along
$\mu$, $\nu$ space-time directions. It is straightforward to calculate plaquette matrix $U_p$
as ordered product of links. For fundamental representation
there are two eigenvectors $\ket{\vec{n}_{\pm}}$: $U_p \ket{\vec{n}_\pm} = e^{\pm i \varphi} \ket{\vec{n}_\pm}$
which are related by $\vec{n}_+ = - \vec{n}_-$. While only a single plaquette
is considered there is no way to distinguish between $\ket{\vec{n}_+}$, $\ket{\vec{n}_-}$
and one can take either of them  as initial state $\ket{\vec{n}_1}$ to be  ascribed to point 1
on the plaquette (see Fig.~1).  Starting from $\ket{\vec{n}_1}$ 
one builds the corresponding evolving states $\ket{\vec{n}_i}$ in all other plaquette corners:
$U_\mu(x) \ket{\vec{n}_1} = e^{i\varphi_1} \ket{\vec{n}_2}$ etc.
Thus the lattice implementation of Eq.~(\ref{phase},\ref{small}) is obtained.

~

\centerline{\psfig{file=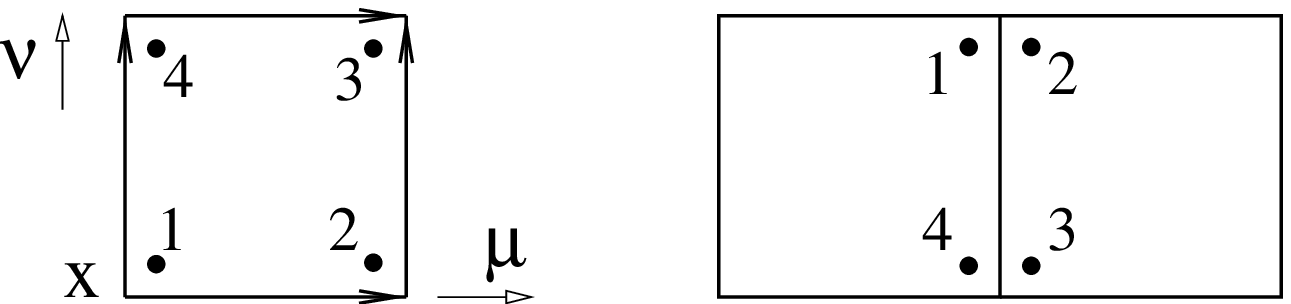,width=0.5\textwidth,silent=}}
\centerline{Fig.~1 ~~~~~~~~~~~~~~~~~~~~~~~~~ Fig.~2}

Consider now the intersection of two plaquettes, Fig.~2. Evidently, on coarse lattice the states
build separately on each plaquette differ drastically. In particular, the states $\ket{\vec{n}_i}$,
$i=1,2$ and $i=3,4$ nearest to common link do not form continuous vector field $\vec{n}$ which is
needed to go from Eq.~(\ref{small}) to Eq.~(\ref{charge}). Moreover, there is also 
a mentioned ambiguity in the choice of initial states $\ket{\vec{n}_\pm}$.
Since the problem is only due to the lattice coarseness
there should be no difference in continuum limit between
various ways to overcome it. We propose to introduce additional two dimensional cells in the intersection
of every two plaquettes which do not lie in the same plane, Fig.~3.
The newly introduced links $V_{12}$, $V_{34}$ may be defined in fact unambiguously. Indeed, among 
various $SU(2)$ matrices which satisfy $V_{12}\ket{\vec{n}_2} = e^{i\alpha}\ket{\vec{n}_1}$ 
there is only one which corresponds to geodesic motion $\vec{n}_2 \to \vec{n}_1$.
Note that these new links are in adjoint representation since for example $\vec{n}_1$ and $\vec{n}_2$
transform in the same

\centerline{\psfig{file=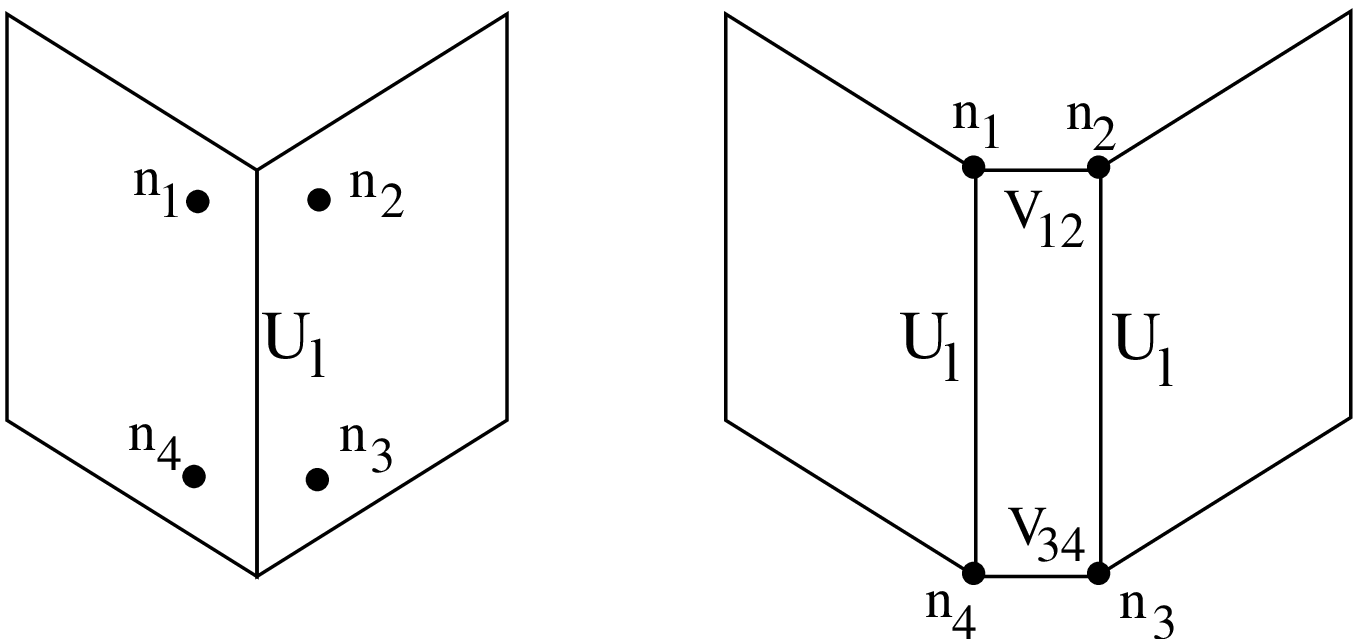,width=0.5\textwidth,silent=}}
\centerline{Fig.~3}

\noindent
way under gauge transformation: $\ket{\vec{n}_i} \to g \ket{\vec{n}_i}$, $i=1,2$.
The ambiguity in choice of initial state on every plaquette
may be resolved in similar manner. Namely, one can impose a requirement that resulting
field $\vec{n}$ should be the smoothest one.

To test the proposed algorithm we have considered the simplest case of single three-dimensional cube. 
The first test was to generate random pure Abelian gauge fields, calculate the monopole charge
in standard way and then apply our method after random non-Abelian gauge transformation.
We found that our method works perfectly on single 3-cube reproducing the known monopole charge in all cases.
Furthermore, we have found a reasonable behavior when non-Abelian gauge fields
$U_l = \mathrm{const} \cdot \{1+ \zeta \,i \Sigma_{i=0}^{3} \varepsilon_i \sigma^i\}$
were generated (here $\varepsilon_i \in [ -1 ; 1 ] $ are random numbers). On Fig.~4 the average
monopole charge as a function of $\zeta$ is shown.

{\bf 3.} We have proposed a gauge invariant definition of Abelian monopole charge in gluodynamics
which is turn is only possible when
gluodynamics is considered as limiting case of lattice gauge
models. The crucial element of the presented construction is the observation that actual symmetry
group of $SU(2)$ LGT is in fact $SU(2)\times U(1)$ and

\centerline{\psfig{file=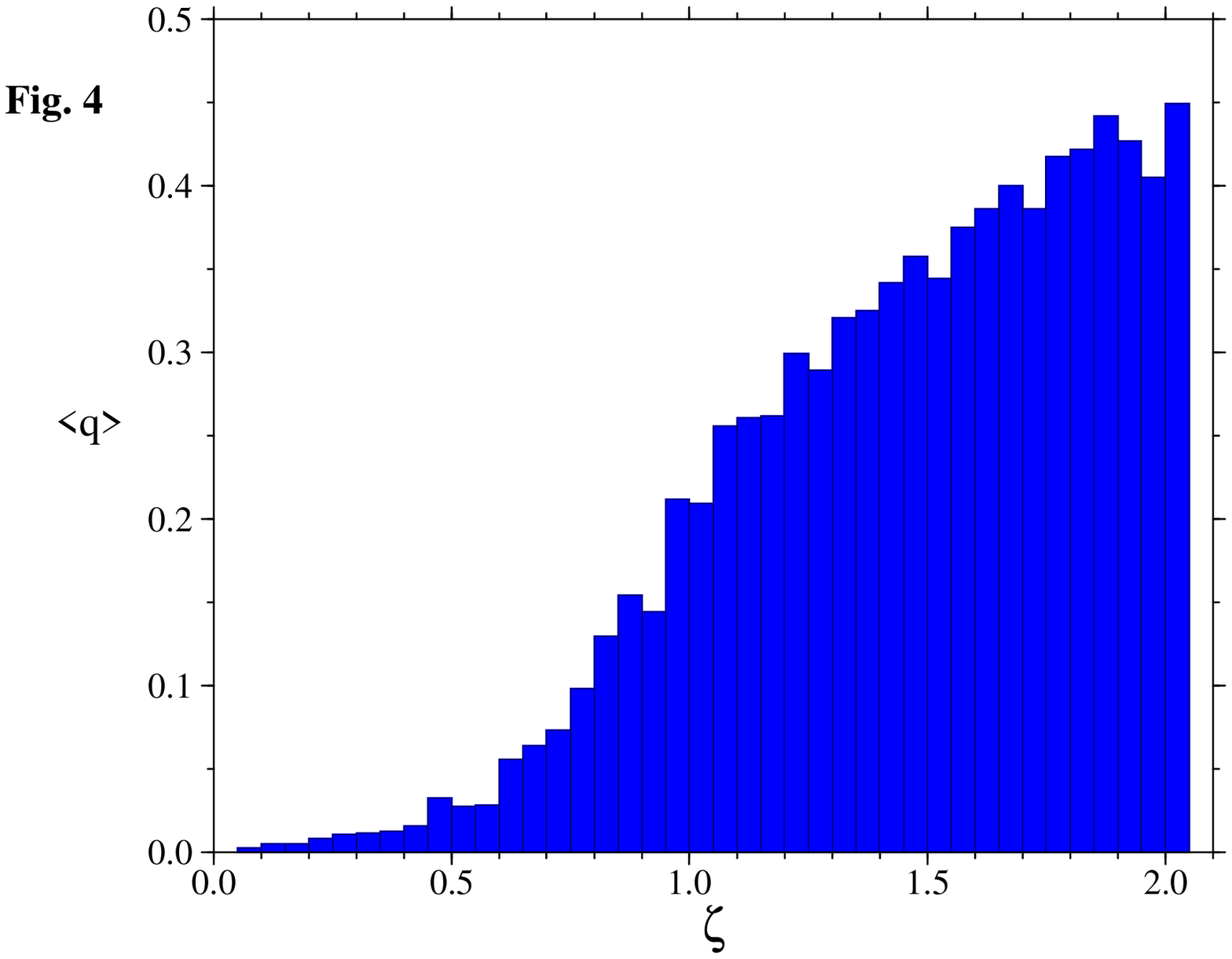,width=0.4\textwidth,silent=}}

\noindent
therefore monopoles defined with respect
to second factor are $SU(2)$ invariant. The explicit calculations are carried out
both in continuum and on the lattice.
Being transparent in continuum limit the actual calculations become rather
intricate on the coarse lattice. We have proposed a way to overcome
this difficulty which allows to investigate the monopole dynamics numerically. Unfortunately,
we were not yet able to implement the method in realistic calculations since 
it requires changing of lattice geometry and standard Monte-Calro algorithms.
The work in these directions is currently in progress.

\end{document}